\documentclass[twocolumn,twoside,slac_two]{revtex4}
\usepackage{graphicx}
\usepackage{fancyhdr}
\begin{document}

\title{V$_{us}$ determination from kaon decays}


\author{P. Massarotti} \thanks{WWW access at www.lnf.infn.it/wg/vus/; for sake of completeness and brevity for all references we refer to the Note written 
by the FlaviaNet Kaon WG~\cite{Flavia2008}.}
\affiliation{Universita' degli studi di Napoli ``Federico
    II'' and INFN Napoli, Naples, Italy}

\begin{abstract}
This review presents the analysis of leptonic and semileptonic kaon decays data
done by the FlaviaNet Kaon Working group, as described 
in~\cite{Flavia2008}.
Data include all recent results by BNL-E865, KLOE, KTeV, ISTRA+, and NA48.
Experimental results are
critically reviewed and combined, taking into account theoretical
(both analytical and numerical) constraints on the semileptonic
kaon form factors. 
We report on a very accurate determination of $V_{us}$ as well as
on many other tests of the SM which 
can be performed with leptonic and semileptonic kaon decays.
\end{abstract}

\maketitle

\thispagestyle{fancy}


\section{Introduction}
\noindent
In the Standard Model, SM, transition rates of semileptonic processes such as 
$d^i \to  u^j \ell \nu$, with $d^i$ ($u^j$) being a generic 
down (up) quark, can be computed with high accuracy in terms 
of the Fermi coupling $G_F$ and the elements $V_{ji}$ of the 
Cabibbo-Kobayashi Maskawa (CKM) matrix. %~\cite{CKM}. 
Measurements of the transition rates provide therefore
precise determinations of the fundamental SM couplings\\A 
detailed analysis of semileptonic decays offers also 
the possibility to set stringent constraints on  new physics scenarios.
While within the SM all $d^i \to  u^j \ell \nu$ transitions 
are ruled by the same CKM coupling $V_{ji}$ (satisfying 
the unitarity condition $\sum_k |V_{ik}|^2 =1$) and 
$G_F$ is the same coupling appearing in the muon decay, 
this is not necessarily true beyond the SM. Setting bounds on the
violations of CKM unitarity, violations of lepton universality, and
deviations from the $V-A$ structure, allows us to put significant
constraints on various new-physics scenarios (or eventually find evidences
of new physics). In the case of leptonic and semileptonic kaon decays these
tests are particularly significant given the large amount of data recently
collected by several experiments: BNL-E865, KLOE, KTeV, ISTRA+, and
NA48. The analysis of these data provides precise determination of
fundamental SM couplings, sets stringent SM test almost free from hadronic
uncertainties, and finally can discriminate between new physics
scenarios. The high statistical precision of measurements  and the detailed
information on kinematical distributions have provided substantial progress
on the theory side, in particular the theoretical error on hadronic form
factors has been reduced at the 1\% level.\\The 
paper is organized as follows. First in Sec.~\ref{BRfits}
we present fits to world data on the leading branching ratios and lifetimes,
for $K_L$, $K_S$, and $K^\pm$ mesons. Sec.~\ref{slopes} summarizes
the status of the knowledge of form factor slopes from $K_{\ell 3}$ decays.
The physics results obtained are described in Sec.~\ref{resulta}, in particular
the measurement of $|V_{us}f_+(0)|$.
\section{Experimental data: BRs and lifetime}
\label{BRfits}
\noindent
Numerous measurements of the principal kaon BRs, or of various ratios
of these BRs, have been published recently. For the purposes of evaluating
$|V_{us}f_+(0)|$, these data can be used in a PDG-like fit to the BRs and
lifetime.
A detailed description to
the fit procedure and the references of all experimental input used 
can be found in Ref.~\cite{Flavia2008}.\\For 
$K_L$ the results are given in table~\ref{tab:KLBR}, while 
table~\ref{tab:KpmBR} gives the results for $K^\pm$.

\begin{table}
\begin{center}
\caption{\label{tab:KLBR} Results of fit to $K_L$ BRs and lifetime. S is
  the scale factor applied to the error in order to obtain $chi^2$ = ndf.}
\begin{tabular}{l|c|r}
Parameter & Value & $S$ \\
\hline
BR($K_{e3}$) & 0.4056(7) & 1.1 \\
BR($K_{\mu3}$) & 0.2705(7) & 1.1 \\
BR(3$\pi^0$) & 0.1951(9) & 1.2 \\
BR($\pi^+\pi^-\pi^0$) & 0.1254(6) & 1.1 \\
BR($\pi^+\pi^-$) & 1.997(7) $\times 10^{-3}$ & 1.1 \\
BR($2\pi^0$) & 8.64(4) $\times 10^{-4}$ & 1.3 \\
BR($\gamma\gamma$) & 5.47(4) $\times 10^{-4}$ & 1.1 \\
$\tau_L$ & 51.17(20)~ns & 1.1 \\
\end{tabular}
\end{center}
\vskip 0.3cm
\end{table}

For the $K_S$, the fit is dominated by the KLOE measurements of
$BR(K_S\to\pi e\nu)$ and of $BR(\pi^+\pi^-)/BR(\pi^0\pi^0)$. These,
together with the constraint that the $K_S$ BRs must add to unity, and the
assumption of universal lepton couplings, completely determine the $K_S$
leading BRs. In particular, BR$(K_S\to\pi e\nu) = 7.046(91) \times 10^{-4}$.
For $\tau_{K_S}$ we use 8.958 $\times 10^{-11}$~s, where this is the non-$CPT$
constrained fit value from the PDG, \cite{PDG06}.

\begin{table}
\begin{center}
\caption{\label{tab:KpmBR} Results of fit to $K^\pm$ BRs and lifetime.}
\begin{tabular}{l|c|r}
Parameter & Value & $S$ \\
\hline
BR($K_{\mu2}$)     & 63.57(11)\%   & 1.1 \\
BR($\pi\pi^0$)      & 20.64(8)\%   & 1.1 \\
BR($\pi\pi\pi$)     &  5.595(31)\%  & 1.0 \\
BR($K_{e3}$)        &  5.078(26)\%    & 1.2 \\
BR($K_{\mu3}$)      &  3.365(27)\%  & 1.7 \\
BR($\pi\pi^0\pi^0$) &  1.750(26)\%  & 1.1 \\
$\tau_\pm$         & 12.384(19)~ns & 1.7 \\
\end{tabular}
\end{center}
\vskip 0.3cm
\end{table}

The fit takes into account the correlation between these values, as
well as their dependence on the $K^\pm$ lifetime.
The world average value for $\tau_\pm$ is nominally
quite precise; the 2006 PDG quotes $\tau_\pm = 12.385(25)$~ns.
However, the error is scaled by 2.1; the confidence level for the
average is 0.17\%. The two new measurements from KLOE ~\cite{KLOE:taupm}
agree with the PDG average, and give a smaller scale factor to the
$\tau_\pm$ value. 
\section{Experimental data: $K_{\ell 3 }$ form factors}
\label{slopes}
\noindent The hadronic  $K \to \pi$ matrix element of the vector current
is described by two form factors (FFs),  $f_+(t)$ and  $f_0(t)$, defined by
$\langle\pi^{-}\left(  k\right)  |\bar{s}\gamma^{\mu}u|K^{0}\left(
p\right) \rangle = (p+k) ^\mu f_+^{}(t) +(p-k) ^\mu f_-^{}(t)
$ and $f_-^{}(t)=\frac{m^2_K-m^2_\pi}{t}\left(f_0^{}(t)-f_+^{}(t)\right)$
where $t=(p-k)^2$.\\ By construction,  $f_0(0)=f_+(0)$.
In order to compute the phase space integrals
we need experimental or theoretical inputs to determinate the  $t$-dependence of FF.
In principle,  Chiral Perturbation Theory (ChPT) 
and Lattice QCD are useful tools to set theoretical constraints.
However, in practice the  $t$-dependence of the FFs at present 
is better determined by measurements and by combining measurements 
and dispersion relations. 
Many approaches have been used, and all have been described in detail
in~\cite{Flavia2008}. For $K_{e3}$ decays, recent measurements of the
quadratic slope parameters of the vector form factor
$({\lambda_+',\lambda_+''})$ are available from KTeV, KLOE, ISTRA+, and
NA48.\\ The same collaborations recentely measured also the slope parameters
$({\lambda_+',\lambda_+'',\lambda_0})$ for $K_{\mu3}$ decays
Here we list only the averages of quadratic fit results for $K_{e3}$ and
$K_{\mu3}$ slopes (\ref{tab:l3ff}) used to determine $|V_{us}|f_+(0)$. It
is important to stress that the significance of the quadratic term in the
vector form factor is strong for both $K_{e3}$ ($4.2\sigma$) and $K_{\mu3}$
($3.6\sigma$) fit to all data. 

\begin{table}
\begin{center}
\caption{Averages of quadratic fit results for $K_{e3}$ and $K_{\mu3}$ slopes.}
\begin{tabular}{l|c}
\hline\hline
                                 & $K_L$ and $K^-$ \\
\hline
Measurements                     & 16		   \\
$\chi^2/{\rm ndf}$               & 54/13 $(7\times 10^{-7})$ \\
$\lambda_+'\times 10^3 $         & $24.9\pm1.1$ ($S=1.4$) \\
$\lambda_+'' \times 10^3 $       & $1.6\pm0.5$ ($S=1.3$)  \\
$\lambda_0\times 10^3 $          & $13.4\pm1.2$ ($S=1.9$) \\
$\rho(\lambda_+',\lambda_+'')$   & $-0.94$                \\
$\rho(\lambda_+',\lambda_0)$     & $+0.33$                \\
$\rho(\lambda_+'',\lambda_0)$    & $-0.44$                \\
$I(K^0_{e3})$                    & 0.15457(29)	      \\
$I(K^\pm_{e3})$                  & 0.15892(30)	      \\
$I(K^0_{\mu3})$                  & 0.10212(31)	      \\
$I(K^\pm_{\mu3})$                & 0.10507(32)	      \\
$\rho(I_{e3},I_{\mu3})$          & $+0.63$             \\
\hline\hline
\end{tabular}
\end{center}
\label{tab:l3ff}
\end{table}

\section{Physics results}
\label{resulta}
\subsection{Determination of $| V_{us}|f_{+}(0)$ and
 $| V_{us}|/| V_{ud}|f_K/f_\pi$}
\noindent
The value of $|V_{us}|f_{+}(0)$ has been determined from 
the decay rate of kaon semileptonic decays: 
\begin{equation}
\label{eq:Mkl3}
\Gamma(K_{\ell 3(\gamma)}) = { G_\mu^2 M_K^5 \over 192 \pi^3} C_K S_{\rm
  ew}\,|V_{us}|^2 f_+(0)^2 \times
\end{equation}
$$
I_K^\ell(\lambda_{+,0})\,\left(1 + \delta^{K}_{SU(2)}+\delta^{K \ell}_{\rm
  em}\right)^2 
$$
using the world average values reported in previous sections
for lifetimes, branching ratios and phase space integrals and the radiative
and $SU(2)$ breaking corrections discussed in~\cite{Flavia2008}.
\begin{figure}[t]
\centering
\includegraphics[width=0.7\linewidth]{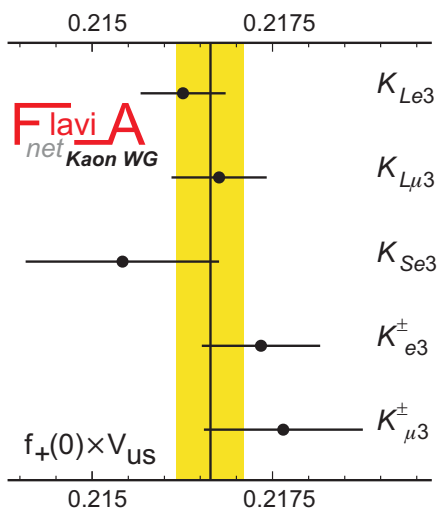}
\caption{ Display of $|V_{us}|f_{+}(0) $ for all channels. 
\label{fig:Vusf0} }
\end{figure}
\begin{table}[t]
\begin{center}
\begin{tabular}{l|c|c|c|c|c|c}
mode               & $|V_{us}|f_{+}(0)$  & \% err & BR   & $\tau$ & $\Delta$& Int \\
\hline
$K_L \to \pi e \nu$     & 0.2163(6)   & 0.28   & 0.09 & 0.19   & 0.15  & 0.09\\
$K_L \to \pi \mu \nu$   & 0.2168(7)   & 0.31   & 0.10 & 0.18   & 0.15  & 0.15\\
$K_S \to \pi e \nu$     & 0.2154(13)  & 0.67   & 0.65 & 0.03   & 0.15  & 0.09\\
$K^\pm \to \pi e \nu$   & 0.2173(8)   & 0.39   & 0.26 & 0.09   & 0.26  & 0.09\\
$K^\pm \to \pi \mu \nu$ & 0.2176(11)  & 0.51   & 0.40 & 0.09   & 0.26  & 0.15\\
 average & 0.2166(5)  &    &  &   &   & 
\end{tabular}
\end{center}
\caption{\label{tab:Vusf0}
 Summary of $|V_{us}|f_{+}(0)$ determination from all channels.}
\end{table}
The results  are shown in figure~\ref{fig:Vusf0} and given in
Table~\ref{tab:Vusf0}, for  $K_L\to\pi e\nu$, $K_L\to\pi\mu\nu$,
$K_S\to\pi e\nu$, $K^\pm\to\pi e\nu$, $K^\pm\to\pi\mu\nu$, 
and for the combination.
The average,   
$|V_{us}|f_+(0)=0.21664(48)$, has an uncertainty of about of $0.2\%$.  
The results from the five modes are in good agreement, the
fit probability is 58\%. 
In particular, comparing the values of $|V_{us}|f_{+}(0)$
obtained from $K^0_{\ell3}$ and $K^\pm_{\ell3}$ we obtain
a value of the SU(2) breaking correction 
$\delta^K_{SU(2)_{exp.}}=2.9(4)\%$
in agreement with the CHPT calculation $\delta^K_{SU(2)}= 2.36(22)\%$.
Moreover, recent analyzes
on the so-called violations of Dashen's theorem in the kaon 
electromagnetic mass splitting point to $\delta^{K}_{SU(2)}$ 
values of about $3\%$.\\ The 
test of  Lepton Flavor Universality (LFU) between  
$K_{e3}$ and $K_{\mu3}$ modes constraints a possible
anomalous lepton-flavor dependence in the leading 
weak vector current. It can therefore be compared
to similar tests in $\tau$ decays, but is different from the 
LFU tests in the helicity-suppressed modes $\pi_{l2}$ and $K_{l2}$.
The results on the parameter
$r_{\mu e} = R_{K_{\mu3}/K_{e3}}^{\rm{Exp}}/R_{K_{\mu3}/K_{e3}}^{\rm{SM}}$ is
$r_{\mu e} = 1.0043 \pm 0.0052$,
in excellent agreement with lepton universality. 
With a precision of $0.5\%$ the test in $K_{l3}$ decays 
has now reached the sensitivity of other determinations:
$r_{\mu e}(\tau) = 1.0005 \pm 0.0041$ and
$r_{\mu e}(\pi) = 1.0042 \pm 0.0033$~\cite{PDG06}\\ An 
independent  determination of $V_{us}$ is obtained from $K_{\ell2}$ decays. 
The most important mode is  $K^+\to\mu^+\nu$, which has been recently 
updated by KLOE reaching a relative uncertainty of about $0.3\%$.  
Hadronic uncertainties are minimized considering the ratio 
$\Gamma(K^+\to\mu^+\nu)/\Gamma(\pi^+\to\mu^+\nu)$:
$$
\frac{\Gamma(K^{\pm}_{\ell 2(\gamma)})}{\Gamma(\pi^{\pm}_{\ell 2(\gamma)})} =
\left|\frac{V_{us}}{V_{ud}} \right|^2\frac{f^2_K m_K}
{f^2_\pi m_\pi}\left(\frac{1-m^2_\ell/m_K^2}{1-m^2_\ell/m_\pi^2}\right)^2
\times\left(1+\delta_{\rm em}\right)
$$
Using the world average values
of BR($K^\pm\to\mu^\pm\nu$) and of $\tau^\pm$ given in Section~\ref{BRfits}
and the value of $\Gamma(\pi^\pm\to\mu^\pm\nu)=38.408(7)~\mu s^{-1}$
from~\cite{PDG06} we obtain:
$|V_{us}|/|V_{ud}|f_K/f_\pi = 0.2760 \pm  0.0006$.

\subsection{Theoretical estimates of $f_+(0)$ and $f_K/f_\pi$ }
\noindent
The main obstacle in transforming these highly precise determinations of 
$|V_{us}|f_{+}(0)$ and 
$|V_{us}|/|V_{ud}|f_K/f_\pi$ into a determination of 
$|V_{us}|$ at the precision of 0.1$\%$, are the theoretical 
uncertainties on the hadronic parameters $f_+(0)$ and $f_K/f_\pi$.
By construction, $f_+(0)$  is defined in the absence of isospin-breaking
effects of both electromagnetic and quark-mass origin.
More explicitly $f_{+}(0)$ is defined 
by the $K^0 \to \pi^+$ matrix element of the vector current in the limit 
$m_u = m_d$ and $\alpha_{\rm em} \to 0$, keeping kaon and pion masses 
to their physical values.
This hadronic quantity cannot be computed in perturbative QCD, but
it is highly constrained by $SU(3)$ and chiral symmetry. 
In the chiral limit and, more generally, in the $SU(3)$ limit
($m_u=m_d=m_s$) the conservation of the vector current
implies $f_+(0)$=1. Expanding around the chiral limit in powers 
of light quark masses we can write
$f_+(0)= 1 + f_2 + f_4 + \ldots$
where $f_2$ and $f_4$ are the NLO and 
NNLO corrections in ChPT.  The Ademollo--Gatto theorem implies that
$(f_+(0)-1)$ is at least of second order in the breaking of $SU(3)$ 
%%%%or in the expansion in powers of $m_s-\hat m$, where $\hat m=(m_u+m_d)/2$. 
This in turn implies 
that  $f_2$ is free from the uncertainties  of the $\mathcal{O}(p^4)$ counter-terms in ChPT, 
and it  can be computed with high accuracy: $f_2=-0.023$. %~\cite{LR}. 
The difficulties in 
estimating $f_+(0)$ begin with $f_4$ or at $\mathcal{O}(p^6)$ in the chiral expansion.
Several analytical approaches to determine $f_4$
have been attempted over the years, %~\cite{f0}, 
essentially confirming the original estimate by Leutwyler and Roos.
The benefit of these new results, obtained using 
more sophisticated techniques, lies in the fact that a 
better control over the systematic uncertainties of the calculation
has been obtained. However, the size of the error is still around or 
above $1\%$, which is not comparable  to the $0.2\%$
accuracy which has been reached for $|V_{us}|f_+(0)$.\\ Recent 
progress in lattice QCD gives us more optimism in the reduction of 
the error on $f_+(0)$ below the $1\%$ level. %~\cite{kan,jut,f0latt}.  
Most of the currently available 
lattice QCD  results have been obtained with relatively heavy pions and  
the chiral extrapolation represents the dominant source of uncertainty.
There is a general trend of lattice QCD results to be slightly lower than
analytical approaches. An important step in the reduction of the error associated to the chiral extrapolation  has been recently made by 
the UKQCD-RBC collaboration. %~\cite{rbcf0}. 
Their preliminary  result $f_+(0)=0.964(5)$
is obtained from the unquenched study  with 
$N_F=2+1$ flavors, with an action that has good chiral properties 
on the lattice even at finite lattice 
spacing (domain-wall quarks). They also reached pions masses ($\geq 330$ MeV) 
much lighter than that used in  previous studies of $f_+(0)$. The 
overall error is estimated to be  $~0.5\%$, which is very encouraging.\\In 
contrast to the semileptonic vector form factor, the pseudoscalar
decay constants are not protected by the Ademollo--Gatto theorem and 
receive corrections linear in the quark masses. Expanding 
$f_K/f_\pi$ in power of quark masses, in analogy to $f_+(0)$, 
$f_K/f_\pi= 1 + r_2 + \ldots$
one finds that the $\mathcal{O}(p^4)$ contribution $r_2$ is 
already affected by local contributions and cannot be unambiguously 
predicted in ChPT. As a result, in the determination of $f_K/f_\pi$
lattice QCD has essentially no competition from purely analytical approaches. 
The  present overall accuracy is about $1\%$. 
The  novelty are the new  lattice results with 
$N_F=2+1$ dynamical quarks  and  pions as light as  $280$~MeV, 
obtained by using the so-called staggered quarks.
These analyzes cover a broad range of lattice spacings (i.e.~$a$=0.06 and
0.15 fm) and are performed on sufficiently large physical volumes ($m_\pi
L\geq 5.0$). It should be stressed, however, that the sensitivity of 
$f_K/f_\pi$ to lighter pions is larger 
than in the computation of $f_+(0)$ and that  chiral extrapolations are far
more demanding in this case.
In the following analysis we will use as reference value the MILC-HPQCD
result $f_K/f_\pi=1.189(7)$. 

\subsection{Test of CKM unitarity}
\begin{figure}[t]
\centering
\includegraphics[width=0.7\linewidth]{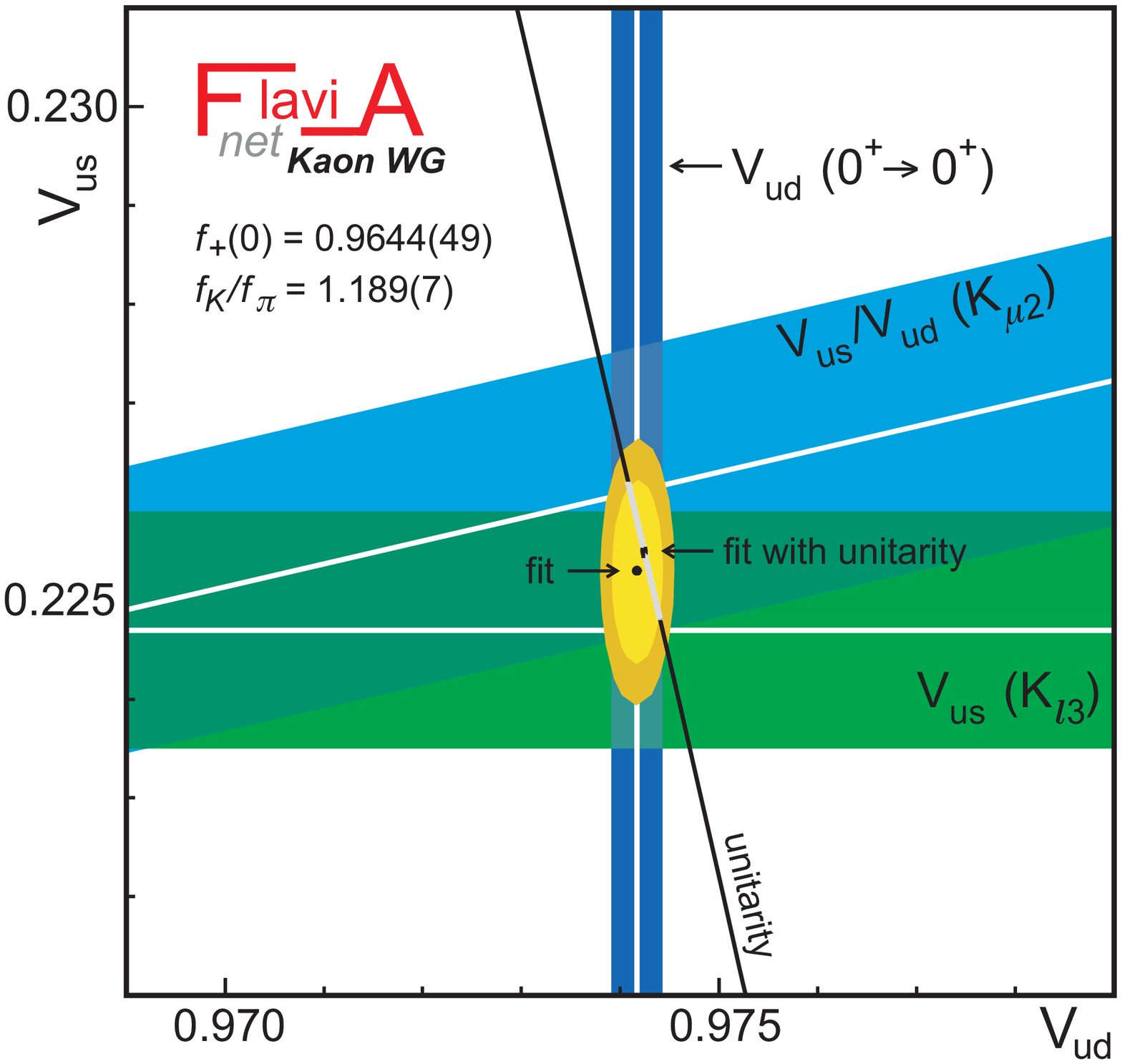}
\caption{\label{fig:vusuni} Results of fits to $|V_{ud}|$, $|V_{ux}|$, and $|V_{us}|/|V_{ud}|$.}
\end{figure}

\noindent
To determine $|V_{us}|$ and   $|V_{ud}|$
 we use the value $|V_{us}| f_{+}(0)=0.2166(5)$, the result $|V_{us}|/|V_{ud}|f_K/f_\pi = 0.2760(6)$, $f_+(0) = 0.964(5)$, and $f_K/f_\pi = 1.189(7)$. 
From the above we find:
$|V_{us}|=  0.2246\pm  0.0012$ from $K_{\ell 3}$ only, and
$|V_{us}|/|V_{ud}|= 0.2321\pm  0.0015$ from $K_{\ell 2}$ only.
These determinations can be used in a fit together with the 
the recent evaluation of $V_{ud}$ from
$0^+\to0^+$ nuclear beta decays: $|V_{ud}|$=0.97418$\pm$0.00026. %~\cite{t&h}.
This global fit gives $V_{ud} = 0.97417(26)$ and $V_{us} = 0.2253(9)$,
with $\chi^2/{\rm ndf} = 0.65/1$ (42\%). This result does not make use 
of CKM unitarity. If the  unitarity constraint is included, 
the fit gives $V_{us}=0.2255(7)$ and $\chi^2/{\rm ndf}=0.80/2$ (67\%).
Both results are illustrated in Figure \ref{fig:vusuni}.
The test of CKM unitarity can be also interpreted as a test of universality of
the lepton and quark gauge couplings.
Using the results of the fit (without imposing unitarity) we obtain:
$G_{\rm CKM} \equiv G_\mu \left[ |V_{ud}|^2+|V_{us}|^2+|V_{ub}|^2 \right]^{1/2}
= (1.1662 \pm  0.0004)\times 10^{-5}\  {\rm GeV}^{-2}$,
in perfect agreement with the value obtained from the measurement
of the muon lifetime:
$G_{\mu} = (1.166371 \pm  0.000007)\times 10^{-5}\  {\rm GeV}^{-2}.$
The current accuracy of the lepton-quark universality
sets important constraints on model building beyond the SM.
For example, the presence of  a $Z^\prime$ would affect the relation between 
$G_{\rm CKM}$ and $G_{\mu}$. In case of a $Z^\prime$ from $SO(10)$ grand unification theories
we obtain  $m_{Z^\prime}>700$~GeV at 95\% CL, to be compared with the  $m_{Z^\prime}>720$~GeV
bound  set through the direct collider searches~\cite{PDG06}.
In a similar way, the unitarity constraint also provides useful bounds in various 
supersymmetry-breaking scenarios. %~\cite{barb}.

\subsection{$K_{\ell 2}$ sensitivity to new physics}
\begin{figure}[t]
\centering
\includegraphics[width=0.7\linewidth]{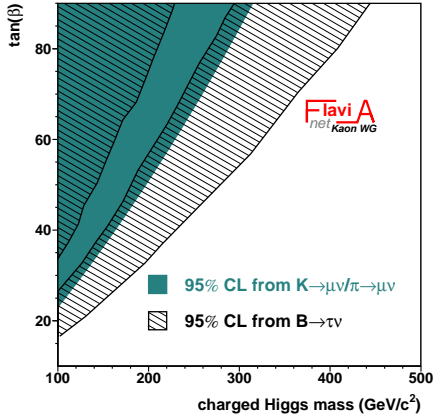}
\caption{\label{fig:higgskmunu}
Excluded region in the charged Higgs mass-$\tan\beta$ plane.
The region excluded by  $B\to \tau \nu $ is also indicated.}
\end{figure}

\noindent
%%\subsection{Bounds on helicity-suppressed amplitudes}
A particularly interesting test is the comparison of the $|V_{us}|$ 
value extracted from the helicity-suppressed $K_{\ell 2}$ decays
with respect to the value extracted from the  helicity-allowed $K_{\ell 3}$  modes.
To reduce theoretical uncertainties from $f_K$ and electromagnetic 
corrections in $K_{\ell 2}$, we exploit the ratio $BR(K_{\ell2})/BR(\pi_{\ell2})$ and 
we study the quantity
$$
R_{l23}=\left|\frac{V_{us}(K_{\ell 2})}{V_{us}(K_{\ell 3})}
\frac{V_{ud}(0^+\to 0^+)}{V_{ud}(\pi_{\ell 2})}\right|\,.
$$
Within the SM, $R_{l23}=1$, while deviation from 1 can be induced by 
non-vanishing scalar- or  right-handed currents.
Notice that in $R_{l23}$ the  hadronic uncertainties enter through  $(f_K/f_\pi)/f_+(0)$.
In the case of effect of scalar currents due to a charged Higgs, 
the unitarity relation between 
$|V_{ud}|$ extracted from $0^+\to0^+$ nuclear beta decays and $|V_{us}|$ extracted from 
$K_{\ell3}$ remains valid as soon as form factors are experimentally determined.
This constrain  together with the experimental information of $\log C^{MSSM}$ 
can be used in the global fit to improve the accuracy of the determination 
of $R_{l23}$, which in this scenario turns to be 
$\left. R_{l23} \right|^{\rm exp}_{\rm scalar} =  1.004 \pm  0.007$.
Here $(f_K/f_\pi)/f_+(0)$ has been fixed from lattice. This ratio
is the key quantity to be improved in order to reduce 
present uncertainty on $R_{l23}$. 
This  measurement of $R_{l23}$ can be used to set bounds
on the charged Higgs mass and $\tan\beta$.
Figure \ref{fig:higgskmunu} shows the excluded region at 95\%
CL in the $M_H$--$\tan\beta$ plane.
The measurement of  BR($B \to \tau \nu$) %~\cite{bib:btaunu}
can be also used to set a similar  bound in the  $M_H$--$\tan\beta$ plane. 
While $B\to\tau \nu$ can exclude quite an extensive region of this plane,
there is an uncovered region in the exclusion corresponding 
to a destructive interference between the charged-Higgs 
and the SM amplitude. This region is fully covered by the $K\to \mu \nu$ result.
%%%%%%%%%%%%%%%%%%%%%%%%%%%%%%%%%%%%%%%%%%%%%%%%%%%%%%%%%

\subsection{A test of lattice calculation}
\noindent
\label{sec:CTtest}

\begin{figure}[t]
\centering
\includegraphics[width=0.7\linewidth]{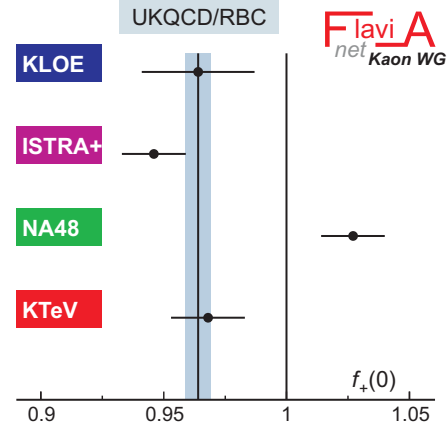}
\caption{Values for $f_+(0)$ determined from the scalar form factor slope using 
the Callan-Treiman relation and  $f_K/f_\pi=1.189(7)$. \label{fig:CTtest} }
\end{figure}

The vector and scalar form factors $f_{+,0}(t)$ are analytic functions in the complex
$t$--plane, except for a cut along the positive real axis, starting at the
first physical threshold $t_{\rm th} = (m_K+m_\pi)^2$, 
where they develop discontinuities. They are real for $t<t_{\rm th}$.
Cauchy's theorem implies that $f_{+,0}(t)$ can be written as 
a dispersive integral along the physical cut where all possible on-shell 
intermediate states contribute to its imaginary part.
A number of subtractions is needed to make the integral convergent.
Particularly appealing is an improved dispersion 
relation recently proposed %in Ref.~\cite{stern} 
where two subtractions are performed at $t=0$
(where by definition, $\tilde f_0(0)\equiv 1$) and at
the so-called Callan-Treiman point $t_{CT} \equiv (m_K^2-m_\pi^2)$.
Since the Callan-Treiman relation fixes the value of scalar form factor at
$t_{CT}$ to the ratio $(f_K/f_\pi)/f_+(0)$,
the dispersive parametrization for the scalar form factor 
allows to transform the available measurements of the scalar form factor 
into a precise information on $(f_K/f_\pi)/f_+(0)$, completely independent of
the lattice estimates. 
Figure \ref{fig:CTtest} shows the values for $f_+(0)$ 
determined from the scalar form factor slope
measurements obtained using a dispersive parametrization and the Callan-Treiman relation, and
$f_K/f_\pi=1.189(7)$. from result on the FF slope using the dispersive 
parameterization
The value of $f_+(0)=0.964(5)$ from UKQCD/RBC is also shown.
The NA48 result is difficult to accommodate. Here one can see that 
this result is also not consistent with the theoretical 
estimates of $f_+(0)$. In particular, it violates the 
Fubini-Furlan bound $f_+(0)<1$. For this reason, the NA48 result will be
excluded when using the Callan-Treiman constraint.

\begin{acknowledgments}
This document is adapted from the instructions provided to the authors
of the proceedings papers at FPCP~06, Vancouver, Canada~\cite{fpcp06},  
and from eConf templates~\cite{templates-ref}.
\end{acknowledgments}

\bigskip % extra skip inserted

\end{document}